\documentstyle[aps,prb,manuscript,psfig]{revtex}
\begin{document}
\draft
\def\dfrac#1#2{{\displaystyle{#1\over#2}}}
\title{$T \geq 0$ properties of the infinitely repulsive Hubbard model 
for arbitrary number of holes.}

\author{P\'eter~Gurin${}^{\rm 1}$ and Zsolt~Gul\'{a}csi${}^{\rm 2}$} 

\address{ 
${}^{\rm 1}$
Institute of Nuclear Research of Hungarian Academy of Sciences, 
H-4026 Debrecen, Hungary.\\
${}^{\rm 2}$ 
Department of Theoretical Physics, Debrecen University,
H-4010 Debrecen, Hungary.}

\date{March, 2000}
 
\maketitle

\begin{abstract} 
Based on representations of the symmetric group $S_N$, explicit and exact 
Schr\"odinger equation is derived for $U \: = \: \infty$ Hubbard model in any
dimensions with arbitrary number of holes, which clearly shows that during the
movement of holes the spin background of electrons plays an important role. 
Starting from it, at $T \: = \: 0$
we have analyzed the behaviour of the system depending on the dimensionality 
and number of holes. Based on the presented formalism thermodynamic quantities
have also been expressed using a loop summation technique in which the 
partition function is given in terms of characters of $S_N$. In 
case of the studied finite systems, the loop summation have been taken into 
account exactly up to the $14$th order in reciprocal temperature and the 
results were corrected in higher order based on Monte Carlo simulations. The 
obtained results suggest that the presented formalism increase the efficiency
of the Monte Carlo simulations as well, because the spin part contribution of
the background is automatically taken into account by the characters of $S_N$.
\end{abstract}

\section{Introduction}

The physics of holes moving in a background build up from a great number of
electrons has received much attention in the last decade given by the
connection of this problem with main questions arising from the study of
strongly correlated systems (Chernyshev and Leung 1999). Developments in
several fields underline this aspect: behaviour in antiferromagnetic
background (Anderson 1997), metal-insulator transition (Trugman 1990 a), doping
effects (Loius {\it et al.} 1999), ferromagnetism (Kollar {\it et al.} 1996),
high $T_c$ superconductivity (Trugman 1998), colossal magnetoresistance
(Ishihara {\it et al.} 1997), spiral states (Watanabe and Miyashita 1999),
generalized-statistics (Long and Zotos 1992), pairing mechanism and
bound-states (Trugman 1990 b), polaronic effects (Louis {\it et al.} 1993), are
main examples to be mentioned. From the theoretical side the Hubbard type
models are especially involved in the theoretical description starting from
infinite on-site repulsion (Nagaoka 1966), finite Hubbard interaction
(S\"utt\H o 1991), and continuing with the inclusion of spin-independent
interactions (Tasaki 1989), extensively extended versions (Kollar {\it et al.}
1996), degenerated orbitals (Shen and Wang 1999) and limiting cases like the
$t-J$ model in one-band (Chernyshev and Leung 1999), or multi-band versions
(Shen and Wang 1999). The theoretical descriptions started with the 1-hole
problem discussed by Nagaoka (Nagaoka 1966), which has been continued with
two-holes studies (Kuz'min 1993), and many-hole analyses in various
configurations and circumstances (Mattis 1986, Doucot and Wen 1989, Barbieri
$et.al.$ 1990, T\'oth 1991, Tian 1991). Despite the huge intellectual effort
spent in the field exact results related to the mentioned problem are
extremely rare (Mattis 1986) and concern only the 1-hole Nagaoka-state
(Nagaoka 1966) and its possible extensions (Tasaki 1989, S\"utt\H o 1991,
T\'oth 1991). However, in the above enumerated concrete applications mostly
multi-hole descriptions are needed mainly in strong-coupling limit. These
situations have been treated in the literature based on various approximations
(acceptable in restricted conditions) like complete separation of charge and
spin degrees of freedom (Kuz'min 1993), extended Gutzwiller wave-functions
(Gebhard and Zotos 1991), canonical transformations (Chernyshev {\it et al.}
cond-mat 9806018), generalization of 1D results to higher dimensions (Doucot
and Wen 1989), restrictions related to the number of holes (Barbieri $et.al.$
1990), unrestricted Hartree-Fock (Loius {\it et al.} 1999), etc., or numerical
procedures like exact diagonalization on small clusters (Takahashi 1982), and
diagonalization within a retained portion of the Hilbert space (Trugman
1998). We witnessed also studies on specific problems like ferromagnetism
(Vollhardt {\it et al.} 1998).

On the background of the mentioned large spectrum of description attempts, in
this paper we are presenting an unifying aspect of the hole-motion in the
simple one-band Hubbard model in the limit of infinitely large on-site
repulsion $U$. For this reason first of all we have deduced and presented an
exact and explicit Schr\"odinger equation for an arbitrary number of holes in
the analyzed model. This equation clearly underlines the role played by the
charge and spin degrees of freedom during the movement of the holes in the $U
= \infty$ Hubbard model, being mathematically based on the representations of
the symmetric group $S_N$ of degree $N$, where $N$ represents the number of
electrons within the system. Advantages of the presented description are
underlined by the observations that: a) the emerging matrix elements can be
find (or deduced in case of larger systems) from standard textbooks of the
group theory, and b) the analysis reduced the concrete equation to be solved
in subspaces fixed by a given, but arbitrary total system spin $S$ of the
initial large Hilbert space, allowing comfortable numerical applications.

After this step, and concrete applications of the
deduced equation related to the $T=0$ spectrum and physical properties of 
small samples, we turn on the study of $T \ne 0$ properties by showing how the
presented formalism helps the deduction of the partition function $Z$ based on
the characters of the representations of the symmetric group (again accessible
from standard text-books in group theory). The emerging coefficients in $Z$
can be reduced to coefficients easily obtainable from
numerical methods, being related to the number of different loops with
fixed length that can be drawn in the studied system connecting its lattice 
sites. The procedure is able even to increase the efficiency of the usual 
Monte Carlo simulation methods. Starting from the deduced formulas, we are 
presenting some examples of the $T \ne 0$ behaviour of the two-hole case.

Our results clearly underline that in the infinitely repulsive
Hubbard model the movement of holes is intimately influenced by the
spin-background of the electrons. As a consequence, in describing this 
process in the general case, the charge and spin degrees of freedom cannot be
entirely separated.  

The remaining part of the paper is organized as follows: Sec. II. describes
the studied model and introduces the used notations, Sec. III. presents the
basis of the Hilbert space used during the paper, Sec. IV. describes the 
action of the Hamiltonian on the basis vectors, Sec. V. presents the deduced 
Schr\"odinger equation together with applications, Sec. VI. concentrates on
$T\ne 0$ properties and emphasises the advantages of the presented formalism 
in deducing thermodynamic quantities like partition function, thermodynamic 
potential, specific heat, expectation value of the total spin, Sec. VII.
presents a summary of the paper, and an Appendix containing the mathematical
details closes the presentation.

\section{The model and notations used}

Our starting Hamiltonian ($\hat H$) describes the one-band Hubbard model in 
$U \: = \: \infty$ limit. Given by the absence of doubly occupied sites in 
the lattice, $\hat H$ becomes
\begin{eqnarray}
{\hat H}_{\infty} \: = \:
\sum_{i,j} \: t_{i j} \: {\hat H}_{\infty}^{i j} \: = \:
\sum_{i,j} t_{i j} \: {\hat P} \: \Bigr( \: \sum_{\sigma} \: 
         {\hat c}_{i, \sigma}^{\dagger} \: {\hat c}_{j, \sigma} \:
\Bigl) \: {\hat P} \: , 
\label{H} 
\end{eqnarray} 
where $t_{i j} \: = \: - \: t$ are hopping matrix elements for
nearest neighbour sites (otherwise $t_{i j} \: = \: 0$). The operators 
${\hat c}_{i, \sigma}^{\dagger} \: ( \: {\hat c}_{i, \sigma} \: )$
creates (annihilates) an electron with spin $\sigma$ at lattice site
$i$, and ${\hat n}_{i} \: = \: \sum_{\sigma} \: {\hat n}_{i, \sigma}$
represents particle number operator. The double occupancy is projected out by 
${\hat P} \: = \: \sum_i \: ( \: 1 \: - \: {\hat n}_{i \uparrow} \: 
{\hat n}_{i \downarrow} \: )$. For convenience, we denote the lattice by 
$\Lambda$, the number of the lattice sites by $N_{\Lambda}$, the number of 
electrons by $N$, so the number of holes becomes
$N_h \: = \: N_{\Lambda} \: - \: N$. Being also interested in magnetic 
properties, we present below the notations and definitions related to spin 
characteristics. The total spin operator is taken as
\begin{eqnarray}
{\hat{\bf S}} \: = \: \frac{1}{2} \: \sum_{i \in \Lambda} \: 
\sum_{\sigma ,\sigma^{\prime}} \: {\hat c}^{\dagger}_{i,\sigma} \: 
\mbox{\boldmath $\tau$}_{\sigma ,\sigma^{\prime}} \: 
{\hat c}_{i,\sigma^{\prime}}
\label{ss}
\end{eqnarray}
where 
$\mbox{\boldmath $\tau$} \: = \: (\tau^x,\tau^y,\tau^z)$ 
are the Pauli matrices. Furthermore we have
${\hat S}^{\pm} \: = \: {\hat S}^x \: \pm \: i \: {\hat S}^y$ and 
${\hat S}^2 \: = \: ( \: {\hat S}^x \: )^2 \: + \: 
( \: {\hat S}^y \: )^2 \: + \: ( \: {\hat S}^z \: )^2$. 
We denote by $S^z$ the eigenvalue of ${\hat S}^z$ and by 
$S \: ( \: S \: + \: 1 \: )$ the eigenvalue of ${\hat S}^2$. 
The quantity $S$ represents the total spin of the state described.

\section{The Hilbert space vector-basis used}

Now we prepare a basis for a sector of the Hilbert space of the system with a
fixed $N_h$ number of holes and a given $S^z$ projection of the total spin. 
We start from Hilbert space vectors characterized by the position of  
holes $h_1  \: , \:  ... \: , \: h_{N_h}$, and the spin configuration of the
electrons situated in single occupied states
$ \{\sigma_i \}_{i \in \Lambda \backslash \{ h_1, ... ,h_{N_h} \} }$, 
defined as
\begin{eqnarray}
|\psi_{\{ h \}, \{ \sigma \} } \rangle =  
\prod_{i \in \Lambda \backslash \{ h_1 , ... , h_{N_h} \}} \: 
{\hat c}^{\dagger}_{i,\sigma_i} \: | \: 0 \: \rangle \: ,
\label{eq1}
\end{eqnarray}
where the product is taken over all the occupied sites in $\Lambda$,
and $|0\rangle$ represents the bare vacuum with no fermions present.
For mathematical convenience, we introduce the convention that 
the elements of $\Lambda$ are building up an ordered set (i.e. the
lattice sites are numbered) and the 
product in Eq.(\ref{eq1}) follows this order. As a consequence, the product
of creation operators from Eq.(\ref{eq1}) is written in increasing order with
respect to lattice indices. Let us introduce now the function
${\cal R} \: : \: \{ \: 1 \: , \: ... \: , \: N_{\Lambda} \: \} \: 
\rightarrow \: \{ \: 1 \: , \: ... \: , \: N_{\Lambda} \: \}$ that performs
a regroupation of lattice site positions, maintaining however their
initial numbering. By definition, ${\cal R}$ gives the indices of occupied 
lattice sites in increasing order for the first $N$ numbers in the
$\{1, ..., N_{\Lambda} \}$ set, and introduces the indices of the
empty sites in arbitrary order for the last $N_h$ positions of the
mentioned set (we underline, $N_{\Lambda} = N + N_h$). The numbering of hole
positions obtained after the action of ${\cal R}$ being arbitrary,
${\cal R}$ itself is not uniquely defined. Using it, 
$|\psi_{\{h\},\{\sigma\}} \rangle$ becomes
\begin{eqnarray}
|\psi_{\{ h \}, \{ \sigma \} } \rangle =  
\prod_{i=1}^N \: {\hat c}^{\dagger}_{{\cal R}(i),\sigma_{{\cal R}(i)}} \: 
| \: 0 \: \rangle \: .
\label{eq2}
\end{eqnarray}

In order to describe the spin configuration with a fixed $S^z$
projection of the total spin, we observe that $S^z$ can be characterized by 
permutations ${\cal P} \: \in \: S_N$ of the symmetric group of degree 
$N$. Indeed, the set $\{ \: \sigma_i \: \}_{i \in \Lambda  
\backslash \{ h_1, ... ,h_{N_h} \} }$ represents in fact a permutation 
of $N_{\uparrow}$ symbols $\uparrow$ and $N_{\downarrow} \: = \: N_{\uparrow}
\: - \: 2 \: S^z $  symbols $\downarrow$ (see Appendix).  
Using ${\cal P}$, the spin configuration can be denoted as 
$(\sigma_i)_{i \in \Lambda \backslash \{ h_1, ... ,h_{N_h} \} } \: = \: 
( \: \sigma( \: {\cal P}( \: i \: ) \: ) \: )_{i=1}^N$, where the function 
$\sigma$ defined by
$ {\sigma}( \: i \: ) \: := \: \uparrow $ if $ i \: = \: 1 \: , \: 
\dots \: , \: N_{\uparrow} $ and
$ {\sigma}( \: i \: ) \: := \: \downarrow $ if $ i \: = \: N_{\uparrow} \: 
+ \: 1 \: , \: \dots \: , \: N $.
With the introduced notations the basis vectors of the Hilbert space with 
fixed number of holes $N_h$ and fixed total spin projection $S^z$ can be 
written as
\begin{eqnarray}
| \: {\cal R \: , \: P} \: \rangle \: = \: \prod_{i=1}^{N} \: 
{\hat c}^{\dagger}_{{\cal R}(i),\sigma({\cal P}(i))} \: 
| \: 0 \: \rangle \: .
\label{basis}
\end{eqnarray}
The subspaces generated by vectors $\{ \: | \: {\cal R \: ,
\: P} \: \rangle \: \}_{{\cal P}\in S_N}$ with different ${\cal R}$ are 
different but isomorphic parts of the whole Hilbert space of the system.
Let us denote by ${\cal H}^{S^z}_s$ the Hilbert space which is isomorphic 
with all subspaces generated by  $\{ \: | \: {\cal R \: ,
\: P} \: \rangle \: \}_{{\cal P}\in  S_N}$ with arbitrary but fixed
${\cal R}$ and $S^z$ (see Appendix). Similarly ${\cal H}_c$ denotes the 
Hilbert space which is isomorphic with all spaces generated by vectors 
$| \: {\cal R \:, \: P} \: \rangle$ with arbitrary but fixed ${\cal P}$
and $S^z$. We mention that independently on $S^z$,  ${\cal H}_c$  represents 
in fact a Hilbert space of a system of hard core spinless particles. 
With this notation the whole Hilbert space of the system ${\cal H}$ is a 
direct sum over $S^z$ sectors ${\cal H} =
\oplus_{S^z} \: ( \: {\cal H}_c \: \otimes{\cal H}_s^{S^z} \: )$.
We underline, that introducing the notation 
${\cal H}_s = \oplus_{S^z}{\cal H}_s^{S^z}$, we have only
${\cal H} \: \cong \: {\cal H}_c \: \otimes \: {\cal H}_s$.  

In the remaining part of the paper we denote an arbitrary element of 
${\cal H}_s^{S^z}$, ${\cal H}_s$ and ${\cal H}_c$ by $\varphi_s^{S^z}$, 
$\varphi_s$ and $\varphi_c$, respectively. Furthermore the basis vectors of 
${\cal H}_c$ will be written as $| \: {\cal R} \: \rangle$, and the basis 
vectors of ${\cal H}_s^{S^z}$ as $| \: {\cal P} \: \rangle$.
With these notations the basis vectors that we have are
$| \: {\cal R \: , \: P} \: \rangle \: = \:
| \: {\cal R} \: \rangle \: \otimes \: | \: {\cal P} \: \rangle$.
The states $| \: {\cal R} \: \rangle$ describe in fact the charge
distribution and $| \: {\cal P} \: \rangle$ the spin
con\-figur\-ation. We have to mention that Eq.(\ref{basis}) has to be used
with care because it contains redundant information. 
Every ${\cal R} \: \in \: S_{N_{\Lambda}}$ 
and ${\cal P} \: \in \: S_N$ uniquely define a
basis vector $| \: {\cal R} \: \rangle \: \otimes \: | \: {\cal P} \:
\rangle$, but several different pairs of permutations define the same
basis vector. Because of this reason indexing these basis vectors by integers,
in the case of the basis set $\{ \: | \: {\cal R}_i \: \rangle \: \} \:
\subset \: {\cal H}_c$ the index $i$ goes up to ${N_{\Lambda} \choose N}$
and not to $N \: !$. Similarly, the basis set $\{ \: | \: {\cal P}_i \: 
\rangle \: \} \: \subset \: {\cal H}^{S^z}_s$ has ${N \choose {N/2 \: + \: 
S^z}}$ different elements. The reason why we express the basis vectors in
term of permutations is detailed in the Appendix: the Hilbert space of
the spin configurations for fixed $N$ and $S^z$ is isomorphic with a
proper right ideal of the group algebra $C [ \: S_N \: ]$
of the symmetric group $S_N$.

We emphasize that the Hamiltonian presented in Eq.(\ref{H})
cannot be written in the product form ${\hat
H}_{\infty} \: \neq \: {\hat H}_c \: \otimes \: {\hat H}_s$ 
and the effect of the Hamiltonian on ${\cal H}_s^{S^z}$ depends on the hole 
configuration (charge distribution) $| \: {\cal R} \: \rangle$. In fact we 
have ${\hat H}_{\infty}^{i j}
\: = \: {\hat H}_c^{i j} \: \otimes \: {\hat H}_s^{i j} \: ( \: {\cal
R} \: )$, therefore the eigenstates of $\hat H$ in general case cannot be 
written in the form $| \: \varphi_c \: \rangle \: \otimes \: |
\: \varphi_s \: \rangle$. As a consequence, there is no $a \: priori$
charge spin separation taken into account at the level of the description.
However, contrary to ${\hat H}$, the effect of spin operators does not
depend on $| \: {\cal R} \: \rangle$, i.e. on ${\cal H}_c$. This is why the 
decomposition of the Hilbert space into sectors characterized by a fixed
total spin $S$ can be done within 
${\cal H}_s \: = \: \oplus_{S^z} \: ( \oplus_{S=S^z}^{S_{max}}
\: {\cal H}_s^{S^z,S})$. Denoting by $\varphi^{S^z,S}$ an element of
${\cal H}^{S^z,S} \: = \: {\cal H}_c \: \otimes \: {\cal H}_s^{S^z,S}$,
an arbitrary common eigenvector of ${\hat S}^z$ and ${\hat S}^2$ in
the sector of fixed number of holes can be written as $\varphi_c \:
\otimes \: \varphi_s^{S^z,S}$.

\section{Acting with the Hamiltonian on basis vectors.}

The contribution ${\hat H}_{\infty}^{i j}$ of the Hamiltonian from Eq.(\ref{H})
has non-vanishing effect on a basis vector from Eq.(\ref{basis}) only if 
$i$ represents a hole position and $j$ an electron position, i.e. 
$ 1 \: \leq \: {\cal R}^{-1}( \: i \: ) \: \leq \: N$ and 
$ N \: + \: 1 \: \leq \: {\cal R}^{-1}( \: j \: ) \: \leq \: N_{\Lambda}$. 
Furthermore, regarding the spin sum in Eq.(\ref{H}), only the term 
${\hat c}_{i, {\sigma}^{\prime}}^{\dagger} \: {\hat c}_{j, {\sigma}^{\prime}}$
of ${\hat H}_{\infty}^{i j}$ has non-vanishing effect in which
${\sigma}^{\prime} \: = \: \sigma( \: {\cal P} \: {\cal R}^{-1}( \: j
\: ) \: )$.  Using the notation $\tilde \jmath \: = \: {\cal R}^{-1}( \: j
\: )$, we obtain 
\begin{eqnarray} 
&&{\hat H}_{\infty}^{i j} \: ( \: | \: {\cal R} \: \rangle \: \otimes \: 
| \: {\cal P} \: \rangle \: ) \: = \: {\hat c}_{i, \sigma({\cal P}({{\tilde 
\jmath}}))}^{\dagger} \: {\hat c}_{j, \sigma({\cal P}({{\tilde \jmath}}))} \: 
\prod_{l=1}^N \: {\hat c}^{\dagger}_{{\cal R}(l), \sigma( {\cal P}(l) )} \: 
| \: 0 \: \rangle \: 
= \: {\hat c}_{i, \sigma({\cal P}({\tilde \jmath}))}^{\dagger} 
\: {\hat c}_{j, \sigma({\cal P}({\tilde \jmath}))} 
\: {\hat c}_{{\cal R}(1), \sigma({\cal P}(1))}^{\dagger}   
\nonumber\\
&& \dots
\: {\hat c}_{{\cal R}({\tilde \jmath}-1), \sigma({\cal P}
({\tilde \jmath}-1))}^{\dagger}
\: {\hat c}_{j, \sigma({\cal P}({\tilde \jmath}))}^{\dagger}  
\: {\hat c}_{{\cal R}({\tilde \jmath}+1), \sigma({\cal P}
({\tilde \jmath}+1))}^{\dagger}
\: \dots \: {\hat c}_{{\cal R}(N), \sigma({\cal P}(N))}^{\dagger}
\: | \: 0 \: \rangle 
\nonumber \\
&&= \: {\hat c}_{{\cal R}(1), \sigma({\cal P}(1))}^{\dagger}  \dots 
\: {\hat c}_{{\cal R}({\tilde \jmath}-1), \sigma({\cal P}
({\tilde \jmath}-1))}^{\dagger}
\: {\hat c}_{i, \sigma({\cal P}({\tilde \jmath}))}^{\dagger}  
\: {\hat c}_{{\cal R}({\tilde \jmath}+1), \sigma({\cal P}
({\tilde \jmath}+1))}^{\dagger} \: \dots 
\: {\hat c}_{{\cal R}(N), \sigma({\cal P}(N))}^{\dagger}
\: | \: 0 \: \rangle 
\nonumber \\
&& = \: \prod_{l=1}^N  \: {\hat c}^{\dagger}_{{\cal P}^{(i,j)} 
{\cal R} (l), \sigma( {\cal P}(l) )} \: | \: 0 \: \rangle 
\label{effect0}
\end{eqnarray}
where ${\cal P}^{(i,j)}$ represents the transposition that interchanges the
indices $i$ and $j$. The result cannot be written in the form $| \: {\cal
P}^{(i,j)} \: {\cal R} \: \rangle \: \otimes \: | \: {\cal P} \:
\rangle$, because the permutation ${\cal P}^{(i,j)} \: {\cal R}$ does
not meet the requirements regarding the site ordering presented after
Eq.(\ref{eq1}). Indeed, the creation operators are no more situated 
within the product in increasing order with respect to their lattice
indices because ${\hat c}^{\dagger}_{i,\sigma({\cal P}({\tilde \jmath}))}$ is
situated in a wrong place.  To overcome this situation, let us 
introduce the cyclic permutation  ${\cal C}(\: {\cal R} \: ; \: i \: , \: j 
\: ) \: \in \: S_N$ which rearranges the $N$ pieces of operators of the last 
row of Eq.(\ref{effect0}) in the required order by moving the 
${\hat c}^{\dagger}_{i,\sigma({\cal P}({\tilde \jmath}))}$ operator 
positioned at site $ {\tilde \jmath} $ to the required position within the
product. In the $i \: < \: j$ $\: (i \: > \: j )$ case, for this purpose 
${\cal C}( \: {\cal R} \: ; \: i \: , \: j \: ) $ interchanges the 
${\hat c}^{\dagger}_{i,\sigma({\cal P}({\tilde \jmath}))}$ operator with 
previous (following) operators. The last operator with which we
interchange, is situated in the position ${\cal R}^{-1}( \:
i^{\prime} \: )$. Here, in the $i \: > \: j$ case  $i^{\prime}$ is the least 
index among the indices greater than $i$ for which 
${\cal R}^{-1}( \: i^{\prime} \: ) \: \leq \: N$, and in the $i \: < \: j$ 
case $i^{\prime}$ is the greatest index among the indices smaller than $i$ 
for which ${\cal R}^{-1}( \: i^{\prime} \: ) \: \leq \: N$. In other words 
$i^{\prime}$ labels the $occupied$ lattice site whose index is closest to $i$
among the indices situated between $i$ and $j$. Now we have
\begin{eqnarray}
{\cal C}( \: {\cal R} \: ; \: i \: , \: j \: ) \: 
= \: {\cal C}^{({\cal R}^{-1}(j) \rightarrow {\cal R}^{-1}(i^{\prime}))}
\: .  
\label{eq3}
\end{eqnarray} 
It can be observed that we get the ordered sequence of the operators 
in Eq.(\ref{effect0}) if we permute their indices by 
${\cal C}^{-1}( \: {\cal R} \: ; \: i \: , \: j \: )$. Given by this,
also a sign-change $( \: - \: 1 \: )^{|{\cal C}({\cal R};i,j)|}$ is
obtained, where $| \: {\cal C} \: |$ has the meaning of the parity of the
permutation. Based on the presented properties we obtain
\begin{eqnarray} \lefteqn{
  {\hat H}_{\infty}^{i j} \: ( \: | \: {\cal R} \: \rangle \: 
  \otimes \: | \: {\cal P} \: \rangle) \:
= \: ( \: - \: 1 \: )^{|{\cal C}({\cal R};i,j)|} \: \prod_{l=1}^N \: 
  {\hat c}^{\dagger}_{ {\cal P}^{(i,j)}{\cal R}{\cal C}^{-1}({\cal R};i,j)(l),
  \sigma( {\cal P} {\cal C}^{-1}({\cal R};i,j)(l) )} \: | \: 0 \: \rangle 
} \qquad
\nonumber \\
& &
= \: | \: {\cal P}^{(i,j)} \: {\cal R} \: {\cal C}^{-1}( 
\: {\cal R} \: ; \: i \: , \: j \: ) \: \rangle \: \otimes \: 
| \: ( \: - \: 1 \: )^{|{\cal C}({\cal R};i,j)|} \: {\cal P} \: 
{\cal C}^{-1}( \: {\cal R} \: ; \: i \: , \: j \: ) \: \rangle \: .
\label{effect1}
\end{eqnarray}

The difference between the hole configuration described by the
permutation ${\cal R}$ and ${\cal P}^{(i,j)} \: {\cal R} \: 
{\cal C}^{-1}( \: {\cal R} \: ; \: i \: , \: j \: )$ lies in the position 
of one hole, which has been moved from site $i$ to site $j$. This shows that
the holes are moved by the Hamiltonian as if they were hard core
bosons, while the spin configuration is changing depending on the
actual hole configuration.  The Hilbert space generated by the vectors
$| \: {\cal R} \: \rangle$ is actually the Hilbert space of $N_h$ hard
core bosons ($| \: {\cal R} \: \rangle$ describes the state in which 
$N_h$ particles are situated in the ${\cal R}( \: N \: + \: 1 \: ) \: ,
\: \dots \: , \: {\cal R}( \: N_{\Lambda} \: )$ lattice sites positions).
Therefore, we write the effect of the Hamiltonian as ${\hat
H}^{i j}_{b,\infty} \: | \: {\cal R} \: \rangle$, taking into account
supplementary the effect on the spin configuration. According to the
Appendix, the effect on the spin configuration can be described with a
linear operator $T^{S^z}[ \: {\cal C}( \: {\cal R} \:
; \: i \: , \: j \: ) \: ]$ where $T^{S^z}$ is a
representation of $S_N$ uniquely determined by $S^z$.  Based on these
considerations, the effect of the Hamiltonian on the basis defined
by Eq.(\ref{basis}) can be given as
\begin{eqnarray}
  {\hat H}_{\infty} \: ( \: | \: {\cal R} \: \rangle \: 
  \otimes \: | \: {\cal P} \: \rangle) \:
= \: \sum_{i,j} \: t_{ij} \: {\hat H}^{i j}_{b,\infty} \: 
  | \: {\cal R} \: \rangle \:  \otimes \: 
  T^{S^z}[ \: {\cal C}( \: {\cal R} \: ; \: i \: , \: j \: ) 
  \: ] \: | \: {\cal P} \: \rangle \: .
\label{effect}
\end{eqnarray}
Starting from Eq.(\ref{effect}) we analyze below the Schr\"odinger equation
describing the behavior of the system.

\section{The Schr\"odinger equation}

Due to the fact that the system is invariant under the SU(2) global
rotations, in the spin space the Hamiltonian and ${\hat S}^2$ are
simultaneously diagonalizable. Therefore the spin eigensubspaces are
invariant under the effect of the Hamiltonian. The effect of the Hamiltonian 
on ${\cal H}_s^{S^z,S}$ can be described by a $T_S$ representation of the 
symmetric group independently of $S^z$ (see Appendix). 
Let us denote by $( \: T_S [ \: {\cal P} \: ] \: )_{nm}$
the matrix elements of the operator $T_S [ \: {\cal P}
\: ]$ in an orthogonal and normalized basis of ${\cal H}_s^{S^z,S}$
with arbitrary $S^z$ (denoted by $\{ \: | b^S_n \rangle \: \}$), and by 
$( \: H^{ij}_{b, \infty} \:)_{kl} \: =
\: \langle \: {\cal R}_k \: | \: {\hat H}^{ij}_{b, \infty} \: | \: 
{\cal R}_l \: \rangle$ where ${\cal R}_k \: , \: {\cal R}_l \: \in \:
\{ \: | \; {\cal R}_i \: \rangle \: \}_{i=1}^{N_{\Lambda} \choose N}$
are basis vectors indexed by integers in $ {\cal H}_c$. Now the
Schr\"odinger equation becomes
\begin{eqnarray}
\sum_{l, m} \: \alpha_{lm} \: \sum_{i, j} \: t_{ij} \:
 ( \: H^{ij}_{b, \infty} \:)_{kl} \: 
 ( \: T_S [ \: {\cal C}( \: {\cal R}_l \: ; i \: , \: j
   \: ) \: ] \: )_{nm} \:
= \: E \: \alpha_{kn}
\label{Schr}
\end{eqnarray}
Eq.(\ref{Schr}) gives the energy eigenvalues $E$ connected to the 
eigenstates with given total spin $S$. The $\alpha_{kn}$ coefficients in
Eq.(\ref{Schr}) represent the coordinates of the energy eigenstate $E$ in the
basis $| \: {\cal R} \: \rangle \: \otimes \: | \: b^S \: \rangle$ given as
\begin{eqnarray}
|\psi_{E,S} \rangle = \sum_{kn} \alpha_{kn} 
| \: {\cal R}_k \: \rangle \: \otimes \: | \: b^S_n \: \rangle
\label{Schrvec}
\end{eqnarray}
We have to mention that essentially the same wave vectors as presented in
Eq.(\ref{Schrvec}) has been used by Long and Zotos (Long and Zotos 1993) in
the study of the stability of the Nagaoka state, without however to write an
explicit Schr\"odinger equation for it.

It can be seen from Eq.(\ref{Schrvec}) that the matrix 
elements of the Hamiltonian between spin states depend also on the hole 
configuration ${\cal R}$, i.e. charge distribution, so
dynamically the spin degrees of freedoms are not separated from the
charge degrees of freedom. At the level of the basis states however, we 
have the possibility to treat them separately and this allows us to write 
the Schr\"odinger equation 
in the form of Eq.(\ref{Schr}). We also can use 
Eq.(\ref{Schr}) for exact numerical diagonalizations.
The main advantage of Eq.(\ref{Schr}) is that the matrix elements of
the starting Hamiltonian have been reduced to a direct sum of 
matrices related to the invariant subspaces ${\cal H}^{S, S^z}_s$.
Because of this reason, in concrete applications we need to solve the 
eigenvalue problem of Eq.(\ref{Schr}) in a relatively small Hilbert space
instead of ${\cal H}^{S^z}_s$ which is much larger.  For concrete 
application the following indications are needed in order to use 
Eq.(\ref{Schr}):
The matrix elements of the hard core boson Hamiltonian can be easily
calculated and for small number of electrons the matrix elements of
the representation $T_S$ is given in mathematical textbooks (see for
example Hamermesh 1962, Table 7.3). For greater $N$ values their
clear calculation procedure is also available from same sources (see
e.g. Hamermesh 1962, Section 7.7).

Eq.(\ref{Schr}) shows explicitly the role played by the total spin $S$ in the
Schr\"odinger equation. For example in the case of $S \: = \: S_{max}$, the
representation $T_{S_{max}}$ is the alternating representation. In this case
the dimension of ${\cal H}_s^{S^z, S_{max}}$ is 1, and the matrix representing
every even permutation is 1 and every odd permutation is $- \: 1$.
Because of this reason hard core fermions with maximal spin do
not differ from spinless fermions (see also Long and Zotos 1993). 
The on-site interaction has no effect on spinless fermions therefore there is
no differention between hard core and free particles in this case.

Another simple example is the 1D case with open boundary conditions.
Numbering the sites of the chain in order one after another, the form of every
nearest neighbour pair is simply $\langle \: i \: , \: i \: + \: 1 \:
\rangle$ in this case, and we have ${\cal C}( \: {\cal R} \: ; \: i \: , \: i 
\: + \: 1 \: ) \: = \: 1 \: \in \: S_N$ whose matrix is the unity matrix for 
arbitrary representation. Therefore there is no difference between hard core 
bosons and fermions, so the energy spectrum is independent of the spin of 
particles, only every energy eigenvalue is $( \: 2 \: S \: + \: 1 \: )^N$ fold
degenerate, $S$ being the spin of particles.

If $D \: > \: 1$ and $S^z \: \neq \: S_{max}$ than the effect of the
Hamiltonian on the spin background is highly nontrivial, and
influences the mobility and interaction of holes. A such type of situation
was analyzed by numerical technique by Trugman (Trugman 1998). He showed that
on N\'eel type antiferromagnetic background one hole is mobile but two
holes are less mobile than it was previously considered. This is due 
in fact in the language presented here by the permutation possibilities of 
spin positions with mobile hole positions. Trugman's results underlined as 
well that the effect of the Hamiltonian is clearly seen also on the spin 
background, which cannot be treated entirely separated from the movement of 
the holes within the system. 

We note that the Schr\"odinger equation Eq.(\ref{Schr})
essentially differs from Kuzmin's equation
(Kuz'min 1993) for the purpose to solve the two hole problem on
singlet background. In the mentioned reference the Schr\"odinger equation 
has been solved supposing implicitly that the Hamiltonian has no effect on
the spin degrees of freedom i.e. spin background. From Eq.(\ref{Schr}) it can
be seen that this is correct in the case of a hard core boson system only. 
This explains why Kuz'min obtains the result that the Nagaoka
state (the fully polarized ferromagnetic state) is degenerated with
the singlet state even in the one hole case, in contradiction with the
Nagaoka's original result (Nagaoka 1966). In one hole case the 
spectrum given by Kuz'min is really related to a hard core 
spinless boson system spectrum, which is equivalent with the spectrum of 
a spinless fermion system. This last one, on its turn, is equivalent 
with the $S_{max}$ spectrum of our original electron system as presented
above. The first equivalence is due to the fact that for hard core spinless 
particles (both bosons and fermions) the $N$ hole problem is equivalent with 
the $N$ particle problem, and in the one particle case the statistics is
unimportant. For more than one hole it is not the same whether we deal with 
bosons or fermions even for hard core interaction. The ground state energy 
of the bosons is always smaller (Long and Zotos 1993). 
In order to exemplify, we mention that the equality presented by Kuz'min
\begin{eqnarray}
0 \: = \: \sum_k \: \frac{1}{E \: - \: ( \: \epsilon_k \: + 
\: \epsilon_{P-k} \: )}
\label{Kuzmin}
\end{eqnarray}
gives the spectrum of a spinless hard core boson system consisting of
two particles (or equivalently two holes). Here $\epsilon_k$ is the one
particle dispersion relation and $P$ is the total momentum of the system.
For the ground state energy of the simplest non-trivial two
dimensional case, the $2 \times 3$ lattice, Eq.(\ref{Kuzmin}) gives
$E_0 \: = \: -6.6468$ while from Eq.(\ref{Schr}) one find 
$E_0 \: = \: -6$ which connected to the $S \: = \: S_{max} \: = \:2$ 
value of the total spin, and $E_{min}^{S=0} \: = \: -5.0212$ is the lowest 
energy value on singlet background. 
Furthermore we note that in the two particle
case the spectrum of hard core spinless boson system is identical with
the spin 1/2 fermion system's singlet spectrum, therefore it can be
computed using Eq.(\ref{Kuzmin}).  However, as presented here, in the $S
\: \neq \: S_{max}$ case the behaviour of the system described by Eq.(\ref{H}) 
cannot be described by hard core particles and independent spin degrees of
freedom. This is the reason why the $N$ particle and the $N$ hole problems 
are not generally equivalent, and the behaviour of few holes is more 
complicated than the behaviour of few particles.

\section{Thermodynamic quantities}

An asset of the representation presented in Eq.(\ref{effect}) is that
the trace of $( \: {\hat H}_{\infty} \: )^l$  over the spin degrees 
of freedom can be computed with this relation exactly. This allows us to
calculate expectation values at $T \ne 0$ of different thermodynamic 
quantities. The procedure is presented as follows.

\subsection{The partition function and total spin}
 
If we are interested in the magnetic properties of the model Eq.(\ref{H}),
we may express for example the expectation value of the square of the total
spin. We have
\begin{eqnarray}
\langle \: {\hat S}^2 \: \rangle \: 
= \: \frac{1}{Z} \: \sum_S \: S \: ( \: S \: + \: 1 \: ) \: 
\sum_{-S \leq S^z \leq S} \: {\rm Tr}_{{\cal H}^{S^z\!\!,S}} \: 
e^{-\frac{{\hat H}_{\infty}}{k T}}
= \: \frac{1}{Z} \: \sum_{S^z} \: 3 \: ( \: S^z \: )^2 \:
{\rm Tr}_{{\cal H}^{S^z}} \: e^{-\frac{{\hat H}_{\infty}}{k T} } \: ,
\label{ter1}
\end{eqnarray}
where the partition function is given by
\begin{eqnarray}
Z \: = \sum_S \sum_{-S \leq S^z \leq S} \:
{\rm Tr}_{{\cal H}^{S^z\!\!,S}} \: e^{-\frac{{\hat H}_{\infty}}{k T}}
\: = \: \sum_{S^z} \: 
{\rm Tr}_{{\cal H}^{S^z}} \: e^{-\frac{{\hat H}_{\infty}}{k T} } \: .
\label{ter2}
\end{eqnarray}
In Eqs.(\ref{ter1},\ref{ter2}) we used the fact that  
${\rm Tr}_{{\cal H}^{S^z,S}} \: e^{-\frac{{\hat H}_{\infty}}{k T} }$
does not depend on $S^z$ due to the SU(2) symmetry, therefore 
$\sum_{S=S^z}^{S_{max}} \: {\rm Tr}_{{\cal H}^{S^z,S}} \: 
e^{-\frac{{\hat H}_{\infty}}{k T} } \: 
= \: {\rm Tr}_{{\cal H}^{S^z}} \: e^{-\frac{{\hat H}_{\infty}}{k T} }$ 
for arbitrary $S^z$. As a consequence, the subspace under the Tr in the last
relations from Eqs.(\ref{ter1},\ref{ter2}) means that the trace should be taken
only over the ${\cal H}^{S^z}$ subspace.

For finite lattices the Tr operation means a finite sum which can be 
interchanged with the sum arising from the series expansion of the 
exponential function. We need in this case for calculations 
${\rm Tr}_{{\cal H}^{S^z}} \: ( \: {\hat H}_{\infty} \: )^l$.
This trace can be computed starting from Eqs.(\ref{basis}, \ref{effect}).
We have
\begin{eqnarray}
&&{\rm Tr}_{{\cal H}^{S^z}} \: ( \: {\hat H}_{\infty} \: )^l \:  
= \: \sum_{|{\cal P}\rangle} \: \sum_{|{\cal R}\rangle} \: 
  \sum_{|{\cal R}_1\rangle} \: \dots \: \sum_{|{\cal R}_{l-1}\rangle} \: 
  \sum_{i_1,j_1} \: \dots \: \sum_{i_l,j_l} \: t_{i_1 j_1} \: \dots \: 
  t_{i_l j_l}  \: \times
\nonumber\\
&&\langle \: {\cal R} \: | \: {\hat H}^{i_l j_l}_{b,\infty} \: 
  | \: {\cal R}_{l-1} \: \rangle \: \dots  \: \langle \: {\cal R}_1 \: | \: 
  {\hat H}^{i_1 j_1}_{b,\infty} \: | \: {\cal R} \: \rangle \: \times  
\nonumber \\
&&\langle \: {\cal P} \: | \: T^{S^z}[ \: {\cal C}( \: 
  {\cal R}_{l-1} \: ; \: i_l \: , \: j_l \: ) \: ] \: \dots \: 
  T^{S^z}[ \: {\cal C}( \: {\cal R}_1 \: ; \: i_2 \: , \: 
  j_2 \: ) \: ] \: T^{S^z}[ \: {\cal C}( \: {\cal R} \: ; \: 
  i_1 \: , \: j_1 \: ) \: ] \: | \: {\cal P} \: \rangle \: .
 \label{tr_H_1}
\end{eqnarray}
Let us denote for $\gamma(\: l \:) \: = \: \gamma(\: 0 \:)$ by 
$\gamma \: \equiv \: ( \: \gamma( \: 0 \: ) \: , \: \gamma( \: 1 \: ) \: , 
\: \dots \: , \: \gamma( \: l \: - \: 1 \: ) \: , \: \gamma( \: l \: ) \: 
) \: = \: ( \: {\cal R \: , \: R}_1 \: , \: \dots \: , \: {\cal R}_{l-1} \: 
, \: {\cal R})$ a sequence of $\cal R$ permutations for which 
$\prod_{i=1}^l \: \langle \: \gamma( \: i \: ) \: | \: {\hat H}_{b,\infty} \: 
| \: \gamma( \: i \: - \: 1 \: ) \: \rangle \: \neq \: 0 $. 
We call $L( \: \gamma \: ) \: := \: l$ the length of $\gamma$. 
The set of all the $\gamma$'s with length $l$ will be denoted by $\Omega_l$.
Let consider now the lattice $\Lambda_h$ consisting of every different 
hole configuration $\cal R$. We recall that $\cal R$ can be described by the
ordered position of holes: $( \: h_1 \: < \: \dots \: < \: h_{N_h} \: )$. 
We consider two lattice points $\cal R$ and $\cal R^{\prime}$ nearest
neighbours if they differ by only one hole position, and these
different hole positions are nearest neighbours in the original
lattice $\Lambda$. The lattice that we obtain in a such a way is a part of 
the $D \: N_h$ dimensional hypercubic lattice, and $\gamma$ is a sequence of
nearest neighbour lattice points in $\Lambda_h$. Because of this reason,
we call $\gamma$ a ,,loop''. In $N_h \: = \: 1$ case $\gamma$ is a loop in 
the original lattice $\Lambda$.

In the case of 
$\sum_{i,j} \: t_{i j} \: \langle \: {\cal R}^{\prime} \: | \: 
{\hat H}_{b,\infty}^{i j} \: | \: {\cal R} \: \rangle \: \neq \: 0 $ 
there is precisely one nearest neighbour pair 
$\langle \: i \: , \: j \: \rangle$ for which 
$\langle \: {\cal R}^{\prime} \: | \: {\hat H}_{b,\infty}^{i j} \: 
| \: {\cal R} \: \rangle \: \neq \: 0$.
Therefore the loop $\gamma$ uniquely determines a sequence of pairs of
nearest neighbour indices $\langle \: i_1 \: , \: j_1 \: \rangle \: ,
\: \dots \: , \: \langle \: i_l \: , \: j_l \: \rangle$ which gives
the only one nonzero contribution of the sums $\sum_{i_1,j_1} \: \dots \:
\sum_{i_l,j_l}$ in Eq.(\ref{tr_H_1}). Denoting the product of
permutations ${\cal C}( \: {\cal R} \: ; \: i \: , \: j \: )$ obtained
from this nonzero term by $\cal P_{\gamma}$, we have 
\begin{eqnarray}
  {\rm Tr}_{{\cal H}^{S^z}} \: ( \: {\hat H}_{\infty} \: )^l \:  
= \: \sum_{\gamma \in \Omega_l} \: \sum_{| {\cal P} \rangle } \: 
  \langle \: {\cal P} \: | \: T^{S^z}[ \: {\cal P}_{\gamma} \: ]
   \: | \: {\cal P} \: \rangle \: = \: \sum_{\gamma \in \Omega_l} \: 
  \chi^{S^z}( \: {\cal P}_{\gamma} \: ).  
\label{ter3}
\end{eqnarray} 
As can be seen from Eq.(\ref{ter3}) we obtain a sum of characters of
$T^{S^z}[ \: {\cal P}_{\gamma} \: ]$. Every character is constant on 
an arbitrary conjugate class $C$. Therefore, the above character sum has 
$N^{(C)}( \: l \: )$ identical members, where $N^{(C)}( \: l \: )$ 
represents the number of the loops with length $l$ for
which ${\cal P}_{\gamma} \: \in \: C$. Using the notation $\beta$ 
for $t$-times the reciprocal temperatures and the results for the sum of the
characters $\chi^{S^z}$ from Eqs.(\ref{sum_of_chi}-\ref{sum2_of_chi}) 
from the Appendix, we obtain
\begin{eqnarray}
  \langle \: {\hat S}^2 \: \rangle \: 
= \: \frac{3}{4 \: Z} \: \sum_{l=0}^{\infty} \: 
  \frac{\beta^l}{l \: !} \sum_{C \subset S_N} \: 
  N^{(C)}( \: l \: ) \: ( \: - \: 1 \: )^{|C|} \: 
  \left( 2^{\sum_{i=1}^N C_i} \: \right) \:
  \left( \: \sum_{i=1}^N \: i^2 \: C_i \: \right)
 \: ,  
\label{S^2}
\end{eqnarray}
where, for the partition function we have 
\begin{eqnarray}
  Z \: 
= \sum_{l=0}^{\infty} \: 
  \frac{\beta^l}{l \: !} \sum_{C \subset S_N} \: 
  N^{(C)}( \: l \: ) \: ( \: - \: 1 \: )^{|C|} \: 
  \left( \: 2^{\sum_{i=1}^N C_i} \right) \: . 
\label{Z}
\end{eqnarray} 
Here $| \: C \: |$ is the parity of permutations from the conjugate
class $C$ and $( \: C_1 \: , \: C_2 \: , \: \dots \: , \: C_N \: )$
describes their cycle structure, i.e. these permutations contains 
$C_i$ cycles with length $i$.

Concerning the technical aspects in using Eq.(\ref{Z}) or Eq.(\ref{S^2}) in 
concrete applications we mention that in order to deduce the coefficients 
$ N^{(C)}( l ) $ one should follow the following procedure. We need 
to deduce a concrete $\sum_{C \subset S_N}$ contribution at a fixed 
$l$ value. For this:
1.) Start from a fixed hole configuration ${\cal R}_1$,
and denote with different numbers every site occupied by electrons.
2.) A single step for a given hole it means that we have to interchange
the given hole with a nearest neighbour number. 
3.) Take $l$ steps with the holes in such a way, that finally get back
the original hole configuration (this is a loop with length $l$). In
this process an arbitrary hole can be moved in every step, the order
of the steps being relevant. However, by interchanging holes between
them, we do not obtain new loops. 
4.) Determine the cycle structure $C$ of the permutation of the
numbers for the obtained loop (i.e. obtain the numbers $C_i$). As a
result of this analyses, we have find in this first step $m=1$ one
loop of length $l$ and a given cycle structure $\{ C_i \}_{m=1}$.
5.) Go back to the step 3). and find a different loop with the same length 
$l$ based on the same starting hole configuration ${\cal R}_1$. After this 
step $m=2$, we have another loop with length $l$, and another cycle structure
$\{ C_i \}_{m=2}$.
6) If you have taken into account every possible different loop starting from
the hole configuration ${\cal R}_1$, do the same procedure with every 
possible different starting hole configuration ${\cal R}_j$.
7) All this being done, the summation $\sum_{C \subset S_N}$ at a fixed $l$
means in fact a summation over all possible cycle structures obtained above. 
For example, given a concrete cycle structure $\{ C_i \}_{m} \: = \:
(C_1 \: = \: a_1, C_2 \: = \: a_2, ... , C_r \: = \: a_r)$ we may express with
it $\sum_i C_i$ or $\sum_i \: i^2 \: C_i$, the quantity $ N^{(C)}( l ) $
having the meaning of how many times this concrete $\{ C_i \}_{m}$ cycle
structure has been obtained in the procedure presented above. The number
$| \: C \: |$ gives the parity of all permutations which give the same
cycle structure $\{ C_i \}_{m}$ (as presented in the Appendix, all different 
permutations with the same cycle structure have the same parity).
8) As presented in the step 7), we have to sum over all possible cycle 
structures obtained in points 1)-6).

Using this procedure, with a computer, a partition function or a spin-square
$T \ne 0$ expectation value can be deduced for a system with arbitrary number
of holes. Evidently, the calculation time increases with the size of the
systems, or number of holes.

\subsection{The free energy and specific heat}

>From the partition function $Z$ given in Eq.(\ref{Z}) we can compute the
free energy $F \: = \: - \: T \: \ln \: Z$ and the specific heat 
$c \: = \: T \: \frac{\partial^2 \: ( \: T \: \ln \: Z \: )}{\partial \: 
T^2}$.

From Eq.(\ref{S^2}) we correctly obtain paramagnetic 
behaviour at high temperatures in thermodynamic limit. The first term
of the $l$-sum becomes dominant if $T$ increases, and $N^{(C)}(
\: 0 \: ) \: = \: 1$ if $C \: = \: 1$ otherwise $N^{(C)}( \: 0 \: ) \:
= \: 0$. Therefore we have in high temperatures limit 
$\langle \: {\hat S}^2 \: \rangle \: = \: 3
\: N \: / \: 4$, so the value of the total spin per particle is
proportional with $ 1 \: / \: \sqrt{N}$.  

In order to analyze the temperature dependence of 
$\langle \: {\hat S}^2 \: \rangle$ we need the values of 
$N^{(C)}( \: l \: )$ as functions of $l$.
Comparing Eq.(\ref{Z}) with the formula for the partition function $Z$
derived in first quantized formalism we obtain
\begin{equation}
\lim_{T \rightarrow 0} \: \frac{\sum_{l=0}^\infty \: 
  \left( \: \frac{t}{k \: T} \: \right)^l \: \frac{1}{l \: !} \: 
  N^{(C)}( \: l \: )}{e^{\frac{- \: E_0}{k \: T}}} \:
= \: \frac{| \: C \: |}{N \: !}
\label{assymptotic}
\end{equation}
for arbitrary conjugate class $C$ which can occur in
Eqs.(\ref{S^2},\ref{Z}), where $E_0$ is the ground state 
energy of the $N$ particle hard core boson system. 

\subsection{The Nagaoka state}

The knowledge of
the asymptotic behaviour given by the Eq.(\ref{assymptotic}) is enough
to analyze the $T$ dependence of the Nagaoka ferromagnetism (i.e. 
$N_h \: = \: 1$ case). In this situation $\Omega_l$ contains the $l$ length
loops of the lattice $\Lambda$. For one hole, $\cal R$ is uniquely
determined by the position of the hole. The positions of the hole 
is described by $\gamma( \: 0 \: )$ at the starting point of the loop, and 
$\gamma( \: l \: ) \: = \: {\cal P}^{i_l,j_l} \:
\dots \: {\cal P}^{i_1,j_1} \: \gamma( \: 0 \: ) \: {\cal
P}_{\gamma}^{-1}$ at the ending point of the loop are the same. 
It can be seen that the
parity of $| \: {\cal P}_{\gamma} \: | \: = \: ( \: - \: 1 \: )^l$ is
always even when we use open boundary conditions, because there is no
loop with odd length in this case.  It means that the dynamical
evolution controlled by the Hamiltonian Eq.(\ref{H}) is not able to permute
the particles by odd permutations, given by the presence of the hard core 
potential acting between them. As a consequence, the fermionic character
of the particles has absolutely no effect in this case. 
This is fully consistent with our previous statement (presented before 
Eq.(\ref{Kuzmin}) ) that the $S=S_{max}$ spectrum of our original fermion 
system is equivalent with the spectrum of a hard core spinless boson system.
Moreover, if bosons ,,existed'' with half spin,
then there would not be differences between the system consisting of
these particles with hard core potential on a square lattice and our
original fermionic system in the one hole case, because the permutations
${\cal C}( \: {\cal R}_l \: ; i \: , \: j \: )$ occur in the 
Schr\"odinger equation Eq.(\ref{Schr}) are always even, and for even 
permutations there are no difference between the representations $T_S$ 
connected to bosons and fermions. As we mentioned above,
the ground state wave function of a boson system is
always symmetric, therefore the spin wave function is also symmetric,
i.e. the ground state is ferromagnetic. This extreme "symmetry" is an 
interesting explanation of Nagaoka's theorem. Our formalism certainly gives
back this result. It is obtained by taking the $T \: \rightarrow \: 0$ limit 
in Eq.(\ref{S^2}), considering the asymptotic behaviour of the
coefficients $N^{(C)}( \: l \: )$ given by Eq.(\ref{assymptotic}) for
the conjugate classes with even parity and $N^{(C)}( \: l \: ) \: = \:
0$ for odd parity, furthermore using the result Eq.(\ref{sum_C_chi})
from the Appendix.
When we use periodic boundary conditions we still have $| \: {\cal
P}_{\gamma} \: | \: = \: ( \: - \: 1 \: )^{L(\gamma)}$ which, however,
can be odd, but only in the case when the linear length of the lattice
is odd at least in one direction, and $\gamma$ goes through the boundary
in this direction. In $t \: < \: 0$ case $\beta^l \: ( \:
- \: 1 \: )^{| C |}$ is always positive, the sign coming from the odd
parity of the permutation is compensated by the sign obtained during
the movement around odd steps, therefore the above argumentation
remains valid.  The key feature now is the fact, that the parity of ${\cal
P}_\gamma$ is in one to one correspondence with the parity of $L( \:
\gamma \: )$.

\subsection{Cases with more than one hole}

The existence of more than one hole permits arbitrary (even and
odd) permutations of the particles under the dynamical evolution
(independently of the length of the loop), therefore the fermionic
feature of them play an important role in building up the energy spectrum of 
the system. Using the asymptotic behaviour of $N^{(C)}( \: l \: )$ from
Eq.(\ref{assymptotic}) and the results of the Appendix presented after
Eq.(\ref{sum_C_chi}), the $\langle \: {\hat S}^2 \: \rangle$ value remains
undetermined (i.e. 0/0), which is due in our interpretation to the fact that 
$E_0$ is not a possible energy for the fermionic system any longer. The 
special similarity between bosons and fermions does not remain valid,
therefore we can not draw conclusion for the magnetic property of the
system based on Nagaoka's theorem. There is a qualitative difference
between the one and two hole case. It is similar with the case when in
a special point the Hamiltonian has an extra symmetry, but the
features of the model could be very different even close to this point
due to the symmetry breaking. If bosons would have half spin, a system 
consisting of such hard core bosons on a lattice would be ferromagnetic even
for arbitrary hole concentration. This statement remains valid for real
bosons with integer (non zero) spin as well and the expectational value of the
total spin goes to a macroscopical (but not to the maximum) value when
the temperature goes to zero, in this case.

In order to obtain information about the two (or more) hole case, the 
study of Eq.(\ref{S^2}) is needed.  To deduce the $N^{(C)}( \: l \: )$
numbers one can use Monte Carlo methods. For example, by random sampling 
we can determine the percentage of the contribution of loops with a given 
length in ${\cal P}_\gamma \: \in \: C$. However, we have to mention, that
the trace over the spin degrees of freedom has been taken already, and 
because of this reason, the presented procedure increases the
efficiency of the usual world line algorithm. In this case, we need only
a sampling from the world lines of the holes, which are $2^N$ times less 
than all the different world lines of the electrons.
To see the trend of the magnetic behaviour in the $T$ dependence, 
we have determined the $N^{(C)}( \: l \: )$ coefficients up to the 14-th 
order exactly, and further, from 15-th up to the 50-th order by Monte Carlo 
in the case of a $5 \: \times \: 5$ lattice with two holes. The square of
the total spin per particle and the specific heat deduced from these
results are presented in Fig1.a. and b. As the temperature decreases
the error of our results increases, but it is clear that the total
spin increases, which underline a tendency to ferromagnetism at $T \to 0$
within the accuracy of our calculation.

\section{Summary}

Summarizing, we have analyzed a simple one-band Hubbard model in $U \: = \:
\infty$ limit in any dimensions. We have treated this problem for the case
in which an arbitrary but fixed hole concentration is present within the
system. Concentrating on the effect of the Hamiltonian on the charge
and spin degrees of freedom and based on the representations of the
symmetric group, we presented an exact Schr\"odinger equation in a form 
which explicitly shows the effect of the spin background in the movement 
of the holes. Explications related to the application of the
presented equation in concrete situations were given in detail. 
Based on the deduced results we have showed that the effect of the spin 
background on the movement of holes is unimportant a) when the system is one 
dimensional with open boundary condition and b) in any dimensions when the 
spin of the background is  maximum. In the first case the holes move as hard 
core spinless bosons, in the second case as hard core spinless
fermions. When we have $D \: > \: 1$ and $S \: \neq \: S_{max}$ the
effect of the spin background seen in the motion of the holes and the
energy spectrum is highly non-trivial. The Nagaoka ferromagnetism 
($N_h \: = \: 1$ case) in our formalism is due 
to the fact that in the one hole case and $U \: = \: \infty$, the
rearrangement of particles with odd parity permutations in the dynamical 
evolution of the system is impossible. Therefore, the fermionic feature is 
not playing any role, and the system is equivalent with
a system of hypothetical hard core half-spin boson system whose ground state 
is trivially ferromagnetic. In the presence of more than one hole this is no 
more true and the complete separation of charge and spin degrees of freedom
is impossible.

On the other hand, based on our representation of the Hamiltonian, the
trace over the spin degrees of freedom can be computed exactly. The presented
procedure allows us to express the partition function, free energy, specific
heat and the expectational value of the square of the total spin. The 
calculations can be given based on the counting of loop contributions, whose
steps were described in extreme details for potential applications. In 
principle the method allows to calculate $T\ne 0$ expectation values for an 
arbitrary number of holes, based on clear formal steps handable by computer. 
In the frame of this formalism the efficiency of the usual Monte Carlo 
methods can also be increased. 

\section*{Appendix} 

In this Appendix we are presenting mathematical details related to deduced
formulas presented in the paper. For the beginning we are presenting an
overview of notations and definitions used, then rather technical proofs 
follow. For more mathematical details connected to the deduction procedure
we refer to Hamermesh 1962. 

We treat a permutation of degree $N$ as a bijective function 
${\cal P} \: : \: \{ \: 1 \: , \: \dots \: , \: N \: \} \: 
\rightarrow \: \{ \: 1 \: , \: \dots \: , \: N \: \}$.
The product of two permutations is defined by the standard composition
of them as functions (from right to left):
$( \: {\cal P} \: {\cal Q} \: )( \: i \: ) \: 
:= \: {\cal P}( \: {\cal Q}( \: i \: ) \: )$. 
The symmetric group of degree $N$ formed by all the permutations of
degree $N$ is denoted by $S_N$. For
$M \: \subset \: \{ \: 1 \: , \: \dots \: , \: N \: \}$ 
the group $S_M$ formed by all the bijections from $M$ onto itself
is a natural subgroup of $S_N$, a function of $S_M$ acts
identically on the elements of 
$\{ \: 1 \: , \: \dots \: , \: N \: \} \: \setminus \: M$.  We
denote the transposition which interchanges $i$ and $j$ by  ${\cal
P}^{(i,j)}$, and ${\cal C}^{(i \rightarrow j)}$ the cycle which moves
$i$ into $j$ while pushing left (right) the elements between $i$ and
$j$ with 1 if $i \: < \: j$ ($i \: > \: j$).  The group algebra 
$C[ \: S_N \: ]$
is a complex Hilbert space generated by the elements of $S_N$ as
an orthonormal basis. The associative but noncommutative product of two
elements of the group algebra defined by the convolution:
\begin{eqnarray*}
  {\sl a} \: {\sl b} \: 
& \equiv & \: \Bigl( \: \sum_{{\cal P} \in S_N} \: \alpha_{\cal P} \: 
  {\cal P} \: \Bigr) \: \Bigl( \: \sum_{{\cal Q} \in S_N} \: 
  \beta_{\cal Q} \: {\cal Q} \: \Bigr) \: 
\nonumber \\
& = & \: \sum_{{\cal P} \in S_N} \: \sum_{{\cal Q} \in S_N} \: 
  \alpha_{\cal P} \: \beta_{\cal Q} \: {\cal P} \: {\cal Q} \: 
= \: \sum_{{\cal P} \in S_N} \:  
  \Bigl( \: \sum_{{\cal Q} \in S_N} \: \alpha_{\cal Q} \: 
  \beta_{{\cal Q}^{-1}{\cal P} } \: \Bigr) \: {\cal P} \: .
\end{eqnarray*}

If a subalgebra $\Im$ has the property that, for ${\sl a}$ in $\Im$,
${\sl a b}$ is also in $\Im$ for any elements ${\sl b}$  of the whole
algebra, then $\Im$ is called a right ideal. If ${\sl e}$ is an
idempotent element of the algebra (${\sl e \: e} \: = \: {\sl e}$) then 
${\sl e} \: C[ \: S_N \: ] \: = \: \{ \: {\sl e \: a} \: 
| \: {\sl a} \: \in \: C[ \: S_N \: ] \: \}$ is
a right ideal, because of the associativity of the group algebra. It
remains true if ${\sl e}$ is only essentially idempotent (${\sl
e \: e} \: = \: c \: {\sl e} \: , \: c \: \in \: C$) because 
${\sl e}^{\prime} \: = \: {\sl e} \: / \: c$ is idempotent.

Now we verify that the Hilbert space generated by all the vectors 
$|{\cal R \: , \: P} \: \rangle$ with arbitrary but fixed ${\cal R}$ 
is isomorphic with a proper right ideal of the group algebra
$C[ \: S_N \: ]$. (In the following $N$ and 
$S^z$ are fixed.) We recall that for fixed $S^z$ we denote by 
$\sigma$ the following  function:
$ {\sigma}( \: i \: ) \: := \: \uparrow $ if 
$ i \: \in \: \{ \: 1 \: , \: \dots \: , \: N_{\uparrow} \: 
= \: N \: / \: 2 \: + \: S^z \: \} $ and
$ {\sigma}( \: i \: ) \: := \: \downarrow $ if 
$ i \: \in \: \{ \: N_{\uparrow} \: + \: 1 \: , \: \dots \: , \: N \: \} $.
Let's denote the subgroup of $S_N$ the elements of which permute
the numbers of the set 
$\sigma^{-1}( \: \uparrow \: )$ by $K_{\uparrow} \: 
= \: S_{\sigma^{-1}( \: \uparrow \: )}$. 
Similarly, the permutations of 
$K_{\downarrow} \: = \: S_{\sigma^{-1}( \: \downarrow \: )}$ 
permute only the $\sigma^{-1}( \: \downarrow \: )$ numbers. Let 
$K \: = \: K_{\uparrow} \: \times K_{\downarrow}$, the direct product of
$K_{\uparrow}$ and $K_{\downarrow}$.  It can be seen that
\begin{eqnarray}
  {\sl e}^{S^z} \: 
:= \: \frac{1}{\sqrt{ \: | \: K \: | \: }} \: \sum_{{\cal P}\in K} \: {\cal P}
\end{eqnarray}
is essentially idempotent, 
($| \: K \: | \: = \: N_{\uparrow} \: ! \: N_{\downarrow} \: !$ is the
order of the subgroup $K$) therefore 
$\Im^{S^z} \: := \: {\sl e}^{S^z} \: C [ \: S_N \: ]$ 
is a right ideal.  For different permutations ${\cal P}$
and ${\cal Q}$ the elements ${\sl e}^{S^z} \: {\cal P}$ and ${\sl
e}^{S^z} \: {\cal Q}$  of the ideal $\Im^{S^z}$ are the same if and only if
the permutations are in the same right coset with respect to the
subgroup $K$ (i.e. ${\cal P} \: = \: {\cal P^{\prime} \: Q},\;\;{\cal
P}^{\prime} \: \in \: K$), because in the expression
$\sum_{{\cal P}\in K} \: {\cal P \: P^{\prime} \: Q} \: = \: 
\sum_{{\cal P}^{\prime\prime}\in K} \: {\cal P^{\prime\prime} \: Q} \: = \: 
\sum_{{\cal Q}^{\prime}\in K{\cal Q}} \: {\cal Q^{\prime}}$ 
the sum is taken over the right coset $K \: {\cal Q}$, and they form a
disjoint cover of the group.

Examining the basis vectors $| \: {\cal R \: , \: P} \: \rangle$ for a
fixed ${\cal R}$ can be seen that the definition Eq.(\ref{basis}) gives
the same vector for different permutations ${\cal P}$ and ${\cal
P^{\prime} \: P}\;\;( \: {\cal P^{\prime}} \: \in \: K \: )$ from the
same right coset, because $\sigma( \: {\cal P}( \: i \: ) \: ) \: = \:
\sigma( \: {\cal P^{\prime} \: P}( \: i \: ) \: ) \: ,\;\; (i \: \in
\: \{ \: 1 \: , \: \dots \: , \: N \: \} \: )$.  However, permutations
${\cal P}_1$ and ${\cal P}_2$ from different right cosets mix the
numbers $\sigma^{-1}( \: \uparrow \: )$ and $\sigma^{-1}( \:
\downarrow \: )$ with each other differently, therefore they lead to
different spin configurations.

The above two paragraphs prove that the map $| \: {\cal R \: , \: P}
\: \rangle \: \rightarrow \: {\sl e}^{S^z} \: {\cal P}$ for fix ${\cal
R}$ is well-defined and injective. Let's extend this map linearly over
the whole space generated by the vectors $| \: {\cal R} \: \rangle \:
\otimes \: | \: {\cal P} \: \rangle$ with fix ${\cal R}$. Certainly we
get a surjective map onto $\Im^{S^z}$, since the elements ${\sl
e}^{S^z} \: {\cal P}$ generate this ideal. Now we show that this
linear isomorphism also preserves the inner product. In the proof we
denote the inner product in the group algebra by $( \: . \: , \: . \: )$.
\begin{eqnarray}
  ( \: {\sl e}^{S^z} \: {\cal P} \: , \: {\sl e}^{S^z} \: {\cal Q} \: ) \: 
& = & \: \Biggl( \: \frac{1}{\sqrt{ \: | \: K \: | \: } } \: 
  \sum_{{\cal S}_1\in K} \: {\cal S}_1 \: {\cal P} \: , \: 
  \frac{1}{\sqrt{ \: | \: K \: | \: }}\sum_{{\cal S}_2 \in K} \: 
  {\cal S}_2 \: {\cal Q} \: \Biggl) \: 
\nonumber \\
& = & \: \frac{1}{| \: K \: |} \: \sum_{{\cal P}^{\prime} \in K{\cal P}} \: 
  \sum_{{\cal Q}^{\prime} \in K{\cal Q}} \: 
  ( \: {\cal P}^{\prime} \: , \: {\cal Q}^{\prime} \: ) \: .
\end{eqnarray}
If the right cosets $K \: {\cal P}$ and $K \: {\cal Q}$ are different
then they have no common elements, therefore $( \: {\cal P}^{\prime}
\: , \: {\cal Q}^{\prime} \: ) \: = \: 0$. (The elements ${\cal P} \:
\in \: S_N$ are orthonormals by definition.) If the right cosets
$K \: {\cal P}$ and $K \: {\cal Q}$ are the same then the sums have $|
\: K \: |$ different members which give 1 (in case ${\cal P}^{\prime}
\: = \: {\cal Q}^{\prime}$), the rest are zero, so the final result is
1.

We showed that the Hilbert spaces generated by the vectors $| \: {\cal
R \: , \: P} \: \rangle$ with arbitrary but fixed ${\cal R}$ are
isomorphic with $\Im^{S^z}$ therefore also with each other. This space
is denoted by ${\cal H}^{S^z}$ instead of $\Im^{S^z}$. The image of $|
\: {\cal R \: , \: P} \: \rangle$ by this isomorphism is independent
of ${\cal R}$. This element of ${\cal H}^{S^z,S}$ is denoted by $| \:
{\cal P} \: \rangle$. Based on the above defined Hilbert space
isomorphism we identify ${\cal H}^{S^z}$ with $\Im^{S^z}$ and we
distinct neither the notations of the inner product nor the operators
act on them anymore. We can regard a linear operator defined on
$\Im^{S^z}$ as an operator on ${\cal H}^{S^z}$ and vice versa.

Similar technical and even more simplier work to prove that the space
${\cal H}_c$ is isomorphic with the Hilbert space of the system of $N$
(or equivalently $N_h$) hard core spinless particles, and its basis
characterized by the positions of these particles uniquely determines
the basis vector $| \: {\cal R} \: \rangle$.

The formula
\begin{eqnarray}
  T[ \: {\cal Q} \: ] \: {\cal P} \: 
:= \: ( \: - \: 1)^{|{\cal Q}|} \: {\cal P \: Q}^{-1}
\end{eqnarray}
defines a representation of the symmetric group in the group algebra.
Here $T[ \: {\cal Q} \: ]$ is a linear operator
associated the ${\cal Q} \: \in \: S_N$ by the representation
$T$, and $\cal P$ is a vector as an element of
$C[ \: S_N \: ]$. This representation is fully
reducible. It is clear from the definition that every right ideal is
an invariant subspace of this representation, thus
$T^{S^z} \: := \: T|_{\Im^{S^z}}$ is also
a representation (with lower dimension) of the symmetric group in
$\Im^{S^z}$.  Comparing the above defined representation and the
effect of the Hamiltonian on a basis vector $| \: \cal R \: \rangle \:
\otimes \: | \: P \: \rangle$ as can be seen from Eq.(\ref{effect1}),
it is clear that the effect on the spin configuration can be described
by the operator $T^{S^z}[ \: {\cal C}( \: {\cal R} \: ;
\: i \: , \: j \: ) \: ]$, and we get Eq.(\ref{effect}).

Now we compute the character of the representation
$T^{S^z}$. By definition the character is the trace of
the operator 
$T^{S^z}[ \: {\cal P} \: ]$ for a ${\cal P} \: \in \: S_N$.
\begin{eqnarray}
\lefteqn{
  \chi^{S^z}( \: {\cal P} \: ) \: 
= \: \sum_{|{\cal Q}\rangle} \: \langle \: {\cal Q} \: | \: 
    T^{S^z}[ \: {\cal P} \: ] \: | \: {\cal Q} \: \rangle \: 
= \: \frac{1}{| \: K \: |} \: \sum_{{\cal Q}\in S_N} \:  
    \langle \: {\sl e}^{S^z} \: {\cal Q} \: | \: 
    T^{S^z}[ \: {\cal P} \: ] \: | \: {\sl e}^{S^z} \: 
    {\cal Q} \: \rangle \: } 
\nonumber \\
& &
= \: \frac{1}{| \: K \: |} \: \sum_{{\cal Q}\in S_N} \: 
    \langle \: {\sl e}^{S^z} \: {\cal Q} \: | \: 
    ( \: - \: 1 \: )^{|{\cal P}|} \: {\sl e}^{S^z} \: 
    {\cal Q \: P}^{-1} \: \rangle \: 
= \: \frac{( \: - \: 1)^{|{\cal P}|}}{| \: K \: |^2} \: 
    \sum_{{\cal Q} \in S_N} \: \sum_{{\cal S}_1,{\cal S}_2 \in K} \: 
    \langle \: {\cal S}_1 \: {\cal Q} \: | \: {\cal S}_2 \: 
    {\cal Q \: P}^{-1} \: \rangle \: 
\nonumber \\
& &
= \: \frac{( \: - \: 1)^{|{\cal P}|}}{| \: K \: |^2} \: 
    \sum_{{\cal Q} \in S_N} \: \sum_{{\cal S}_1,{\cal S}_2 \in K} \: 
    \delta_{{\cal S}_1{\cal Q},{\cal S}_2 {\cal QP}^{-1}} \: 
= \: \frac{( \: - \: 1 \: )^{|{\cal P}|}}{| \: K \: |} \: 
    \sum_{{\cal Q} \in S_N} \: 
    \sum_{{\cal S} \in K} \: \delta_{{\cal QPQ}^{-1},{\cal S}} \: .
\label{character}
\end{eqnarray}
Every element ${\cal P}^{\prime}$ of the conjugate class $C^{\cal P}$
which contains ${\cal P}$ appears as ${\cal Q \: P \: Q}^{-1}$ the
same times.  That's why there are $| \: S_N \: | \: / \: | \:
C^{\cal P} \: |$ permutations ${\cal Q}$ for which ${\cal Q \: P \:
Q}^{-1} \: = \: {\cal P}^{\prime}$. ($| \: X \: |$ means the number of
the elements of the set $X$.) If ${\cal P}^{\prime}$ is in $K$ then
the second sum in the last row of Eq.(\ref{character}) has one
non-zero element, otherwise it does not. Therefore the character
\begin{eqnarray}
  \chi^{S^z}( \: {\cal P} \: ) \: 
&=& \: ( \: - \: 1 \: )^{|{\cal P}|} \: 
    \frac{| \: S_N \: | \: | \: C^{\cal P} \cap K|}
     { \: | \: K \: | \: | \: C^{\cal P} \: | }
\nonumber \\
&=& \: ( \: - \: 1 \: )^{|{\cal P}|} \hspace{-1cm}
  \sum_{\mbox{$ \begin{minipage}[t]{4.4cm}  
      \begin{center} \scriptsize \baselineskip 1mm 
        $c^{\prime}_1, c^{\prime \prime}_1, \dots , 
                 c^{\prime}_N, c^{\prime \prime}_N$ \\  
        $c^{\prime}_1 + c^{\prime \prime}_1 = C^{\cal P}_1, \dots , 
                 c^{\prime}_N + c^{\prime \prime}_N = C^{\cal P}_N $ \\
        $c^{\prime}_1 + 2 c^{\prime}_2 + \dots + N c^{\prime}_N 
                 = N/2 + S^z $ \\
        $c^{\prime \prime}_1 + 2 c^{\prime \prime}_2 + \dots + 
                 N c^{\prime \prime}_N = N/2 - S^z $
      \end{center}
      \end{minipage} $}} \hspace{-.5cm}
  \frac{C^{\cal P}_1 \: !}{c^{\prime}_1 \: ! \: c^{\prime \prime}_1 \: ! } \:
  \: \dots \: \frac{C^{\cal P}_N \: !}{c^{\prime}_N \: ! \: 
  c^{\prime \prime}_N \: ! } 
\nonumber \\
&=& \: ( \: - \: 1 \: )^{|{\cal P}|} \hspace{-.5cm}
  \sum_{\mbox{$ \begin{minipage}[t]{3cm}  
      \begin{center} \scriptsize \baselineskip 1mm 
        $c_1, \dots , c_N $\\  
        $\sum_{i=1}^N i c_i = N/2 + S^z $ 
      \end{center}
      \end{minipage} $}} \hspace{-.5cm}
  \prod_{i=1}^N \: {C^{\cal P}_i \choose c_i} \: ,
\label{character_explicit}
\end{eqnarray}
where the cycle structure of every elements of the conjugate class
$C^{\cal P}$ is described by the $N$ pieces of numbers $C^{\cal P}_1 \: , \:
C^{\cal P}_2 \: , \: \dots \: , \: C^{\cal P}_N$ by the following way:
these permutations contains $C^{\cal P}_1$ cycles with length 1, 
$C^{\cal P}_2$ cycles with length 2, etc. For the second equality 
see Hamermesh 1962, the third can be obtained with simple computation.

We need the sum of these characters on an arbitrary conjugate class
$C \: \subset \: A_N$ where $A_N$ is the alternating group formed by
all the even permutations of $S_N$. Based on the first line of the formula
Eq.(\ref{character_explicit}) we can see, that
\begin{eqnarray}
\sum_{C \subset A_N} \: | \: C \: | \: \chi^{S^z}( \: C \: ) \: 
= \: \frac{N \: !}{| \: K_{S^z} \: |} \: | \: A_N \: \cap \: K^{S^z} \: | \:
= \: \frac{N \: !}{2} \: .
\label{sum_C_chi}
\end{eqnarray}
The similar sum over the conjugate class which has no common element
with $A_N$ gives the result $- \: N \: ! \: / \: 2$. The even
permutations form an invariant subgroup in $S_N$ therefore every
conjugate class is in $A_N$ or disjoint from it. Hence 
$\sum_{C \subset S_N} \: | \: C \: | \: \chi^{S^z}( \: C \: ) \: = \: 0$.

Based on the last line of the formula Eq.(\ref{character_explicit}) we can
get:
\begin{eqnarray}
\sum_{S^z = -N/2}^{N/2} \: \chi^{S^z}( \: C \: ) \:
= \: ( \: - \: 1 \: )^{|C|} \: 2^{\sum_{i=1}^N C_i} \: .
\label{sum_of_chi}
\end{eqnarray}
Furthermore
\begin{eqnarray}
\sum_{S^z = -N/2}^{N/2} \: ( \: S^z \: )^2 \: \chi^{S^z}( \: C \: ) \:
= \: ( \: - \: 1 \: )^{|C|} \: 2^{\sum_{i=1}^N C_i} \:
  \left( \: \frac{1}{4} \: \sum_{i=1}^N \: i ^2 \: C_i \: \right) \: .
\label{sum2_of_chi}
\end{eqnarray}

The representation $T^{S^z}$ is not irreducible if
$S^z \: \neq \: S^z_{max}$ or $S^z_{min}$. Certainly it can be verified in
pure mathematical way, but physically it is due to the SU(2) rotational
symmetry in the spin space. The operator  ${\hat S}^-$ realizes a
linear bijection between the spaces ${\cal H}_s^{S^z,S}$ and ${\cal
H}_s^{S^z-1,S}$ if $S \: \geq \: S^z$. Moreover 
$[ \: {\hat H}_{\infty} \: , \: {\hat S}^- \: ] \: = \: 0$, 
therefore the effect of the Hamiltonian
is equivalent on the space ${\cal H}_s^{S^z,S}$ and ${\cal H}_s^{S^z-1,S}$. 
The effect of the Hamiltonian on the space 
${\cal H}_s^{S^z_{max}} \: = \: {\cal H}_s^{S^z_{max},S_{max}}$ 
can be described by the representation
$T^{S^z_{max}} \: =: \: T_{S_{max}}$, 
hence it is true for the space ${\cal H}_s^{S^z_{max}-1,S_{max}}$. 
However, the effect of the Hamiltonian on ${\cal H}_s^{S^z_{max}-1}$ 
is described by the representation $T^{S^z_{max}-1}$, 
which according to the above coincides with  
$T_{S_{max}}$ on ${\cal H}_s^{S^z_{max}-1,S_{max}}$ 
which is therefore an invariant subspace. Therefore 
$T^{S^z_{max}-1}$ is not irreducible therefore
it is fully reducible (because it is a finite dimensional
representation of a finite group). Therefore the orthogonal
complementer of ${\cal H}_s^{S^z_{max}-1,S_{max}}$ which is the 
${\cal H}_s^{S^z_{max}-1,S_{max}-1}$ is also an invariant subspace. The
restriction of $T^{S^z_{max}-1}$ to this subspace we
denote by
$T_{S_{max}-1} \: := \: T^{S^z_{max}-1}|_{
{\cal H}_s^{S^z_{max}-1,S_{max}-1}}$. 
Continuing this process on this line, acting by ${\hat S}^-$ again and
again, we get:  $T^{S^z}  \: \cong  \: \oplus_{S=|S^z|}^{N/2} \: 
T_{S}$.

In other words this means that the functions $\varphi^{S^z,S}_s$, which
describe the spin configuration with $S$ and $S^z$ values of the total
spin and its $z$ component for fixed hole configuration, form a space of
a representation of the symmetric group. This above defined
representation depending only on the value of $S$ was denoted by
$T_{S}$, and the effect of the Hamiltonian on the spin
configuration can be described by this representation:
\begin{eqnarray}
  {\hat H}^{i j}_{\infty} \: | \: \varphi^{S^z,S}_s \: \rangle \: 
= \: T_{S}[ \: {\cal C}( \: {\cal R} \: ; \: i \: , \: j \: ) 
  \: ] \: | \: {\varphi^{S^z,S}_s} \: \rangle 
\end{eqnarray}

The matrices of the representations $T_S$ can be given
by algebraic methods, see
e.g. Hamermesh 1962. These methods are usable
only in case of few particles. In this case we can use the expression
Eq.(\ref{Schr}) to get exact diagonalization results.

\acknowledgements 
For Zs. G. research supported by the OTKA-T-022874 and FKFP-0471 contracts
of the Hungarian Founds for Scientific research.

\section*{References}

	\mbox{}\\
{\sc Anderson, P. W.,} 
	1997, {\sl Adv. in Phys.,} {\bf 46}, 3.
\\
{\sc Barbieri, A., Riera, J. A.,} and {\sc Young, A. P.,}
	1990, {\sl Phys. Rev. B,} {\bf 41}, 11697.
\\
{\sc Chernyshev, A. L.,} and {\sc Leung, P. W.,}
	1999, {\sl Phys. Rev. B,} {\bf 60}, 1592.
\\
{\sc Chernyshev, A. L., Leung, P. W.,} and {\sc Gooding, R. J.,}
	cond-mat 9806018.
\\
{\sc Doucot, B.,} and {\sc Wen, X. G.,}
	1989, {\sl Phys. Rev. B,} {\bf 40}, 2719.
\\
{\sc Gebhard, F.,} and {\sc Zotos, X.,}
	1991, Phys. Rev. {\bf B43}, 1176,
\\
{\sc Hamermesh, M.,} 1962, 
	Group theory and its application to physical problems, 
	Addison-Wesley Publishing Company, Inc.
\\
{\sc Ishihara, S., Yamanaka, M.,} and {\sc Nagaosa, N.,}
	1997, {\sl Phys. Rev. B,} {\bf 56}, 686.
\\
{\sc Kollar, M., Strack, R.,} and {\sc Vollhardt, D.,}
	1996, {\sl Phys. Rev. B,} {\bf 53}, 9225.
\\
{\sc Kuz'min, E. V.,} 
	1993, {\sl Pis'ma Zh. Eksp. Teor. Fiz.} {\bf 57}, 575.
\\
{\sc Long, M. W.,} and {\sc Zotos, X.,} 
	1992, {\sl Phys. Rev. B} {\bf 45}, 9932; 
	1993, {\sl ibid.,} {\bf 48}, 317.
\\
{\sc Louis, E., Chiappe, G., Guinea, F., Verg\'es, J. A.,} and 
	{\sc Anda, E. V.,} 1993, {\sl Phys. Rev. B,} {\bf 48}, 9581.
\\
{\sc Louis, E., Guinea, F., L\'opez, S. M. P.,} and {\sc Verg\'es, J. A.,}
	1999, {\sl Phys. Rev. B,} {\bf 59}, 14005. 
\\
{\sc Nagaoka, Y.,} 
	1966, {\sl Phys. Rev.} {\bf 147}, 392. 
\\
{\sc Mattis, D. C.,} 
	1986, {\sl Rev. Mod. Phys.,} {\bf 58}, 361. 
\\
{\sc Shen, S. Q.,} and {\sc Wang, Z. D.,} 
	1999, {\sl Phys. Rev. B,} {\bf 59}, 9291.
\\
{\sc S\"utt\H o, A.,} 
	1991, {\sl Commun. Math. Phys.,} {\bf 140}, 43. 
\\
{\sc Takahashi, M.,} 
	1982, {\sl Jour. Phys. Soc. Jpn.,} {\bf 51}, 8475.
\\
{\sc Tasaki, H.,} 
	1989, {\sl Phys. Rev. B,} {\bf 40}, 9192.
\\
{\sc Tian, G. S.,} 
	1991, {\sl Phys. Rev. B,} {\bf 44}, 4444. 
\\
{\sc T\'oth, B.,} 
	1991, {\sl Lett. in Math. Phys.,} {\bf 22}, 321.
\\
{\sc Trugman, S. A.,} 
	1988, {\sl Phys. Rev. B,} {\bf 37}, 1597;
	1990 a, {\sl ibid.,} {\bf 41}, 892; 
	1990 b, {\sl ibid.,} {\bf 42}, 6612.
\\
{\sc Vollhardt, D., Bl\"umer, N., Held, K., Kollar, M., Schlipf, J., 
	Ulmke, M.,} and {\sc Wahle, J.,} 
	1998, Advances in Solid State Physics, Vol. 38.
\\
{\sc Watanabe, Y.,} and {\sc Miyashita, S.,} 
	1999, {\sl Jour. Phys. Soc. Jpn.,} {\bf 68}, 3086.

\begin{figure}
\psfig{file=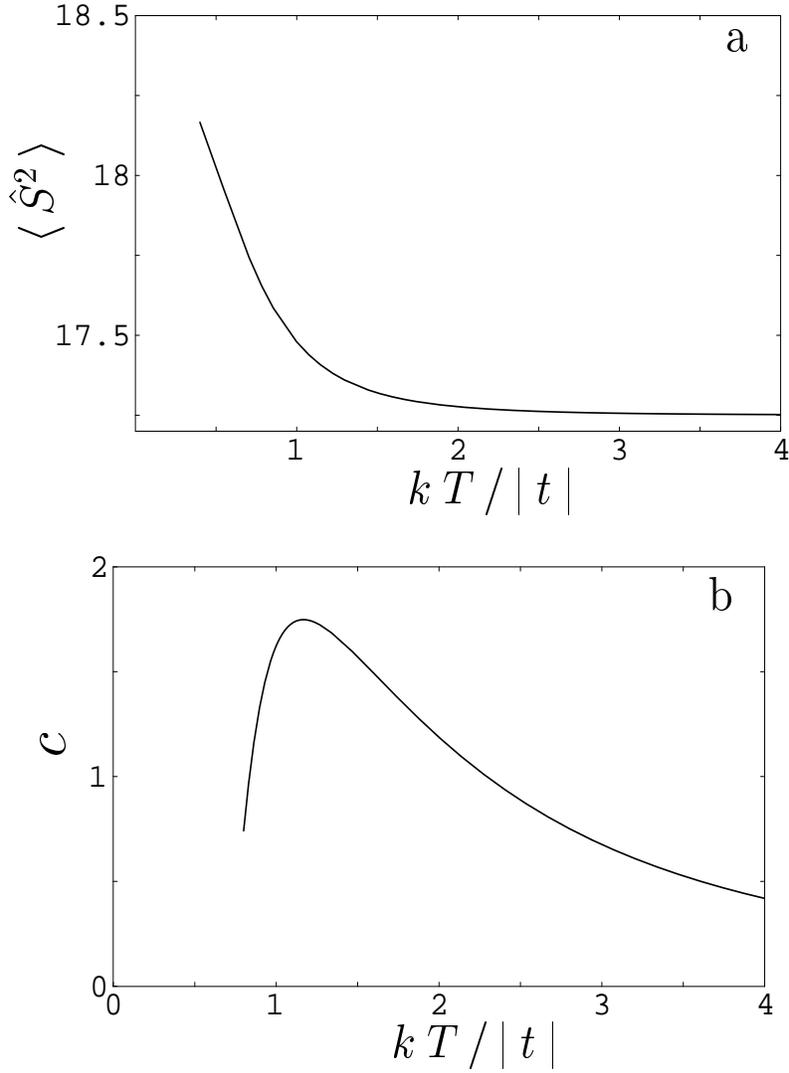,width=12cm,height=16cm}
\caption{
The a: sqare of the total spin and b: specific heat as a function of 
the reduced temperature in the case of $5 \times 5$ lattice with two 
holes. The calculation based on the Eqs.\ref{S^2}-\ref{Z} in which the 
coefficients $ N^{(C)}( l ) $ are taken into account exactly up to 
$ l=14 $ and determined by Monte Carlo simulation up to $ l=50 $.}
\end{figure}

\end{document}